\documentclass[a4paper, 12pt, english]{article}

\usepackage[utf8]{inputenc}
\usepackage{amsmath,amssymb}
\usepackage{graphicx}
\usepackage{subfig}
\usepackage[colorinlistoftodos]{todonotes}

\usepackage{amsmath}
\usepackage{amssymb}
\usepackage{graphicx}
\usepackage{algorithm2e}
\usepackage{breqn}
\usepackage{multirow}
\usepackage{capt-of}
\usepackage[T1]{fontenc}
\usepackage{letltxmacro}
\usepackage{listings}
\usepackage{color}

\definecolor{dkgreen}{rgb}{0,0.6,0}
\definecolor{gray}{rgb}{0.5,0.5,0.5}
\definecolor{mauve}{rgb}{0.58,0,0.82}

\usepackage{indentfirst}
\usepackage{verbatim}
\usepackage{textcomp}
\usepackage{gensymb}

\usepackage{relsize}

\usepackage{lipsum}
\usepackage{xcolor}
\usepackage{xparse}
\NewDocumentCommand{\myrule}{O{1pt} O{2pt} O{black}}{%
  \par\nobreak 
  \kern\the\prevdepth 
  \kern#2 
  {\color{#3}\hrule height #1 width\hsize} 
  \kern#2 
  \nointerlineskip 
}
\usepackage[section]{placeins}

\usepackage{booktabs}
\usepackage{colortbl}%

\usepackage[obeyspaces]{url}
\usepackage{etoolbox}
\usepackage[colorlinks,citecolor=black,urlcolor=blue,bookmarks=false,hypertexnames=true]{hyperref} 

\usepackage{geometry}
\begin{document}

\begin{titlepage}
\begin{center}

\includegraphics[width=0.5\textwidth]{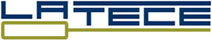}\\
\textbf{\large Laboratoire de Recherches sur les Technologies du Commerce Électronique}\\[0.2cm]

\vspace{10pt}

\includegraphics[width=0.5\textwidth]{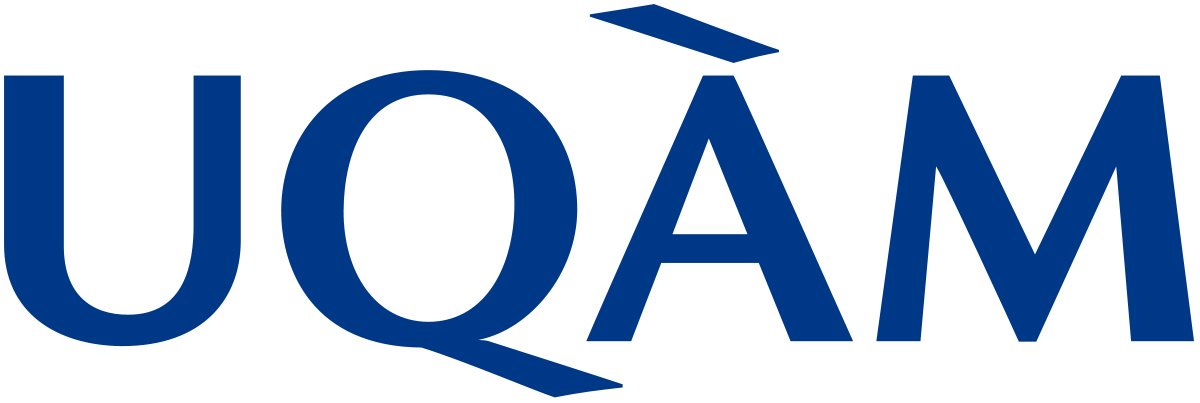}\\[1cm]

\textbf{\LARGE Universit\'e du Qu\'ebec \`a Montr\'eal}\\[0.5cm] 
\vspace{20pt}

\par
\vspace{20pt}
\myrule[1pt][7pt]
\textbf{\Large  How to Implement Dependencies in Server Pages of JEE Web Applications}\\
\myrule[1pt][7pt]

\vspace{35pt}
\large{\textbf{Anas Shatnawi, Hafedh Mili, Manel Abdellatif, \\Ghizlane El Boussaidi, Yann-Ga\"el Gu\'eh\'eneuc, \\Naouel Moha, Jean Privat}}\\[0.3cm]

\textit{LATECE Technical Report 2017-1, LATECE Laboratoire, Universit\'e du Qu\'ebec \`a Montr\'eal, Canada}

\vspace{150pt}
\small{July, 2017}
\end{center}

\par
\vfill
\begin{center}
\end{center}
\end{titlepage}



\newpage

\begin{center}
\textbf{\Large  How to Implement Dependencies in Server Pages of JEE Web Applications} \\[0.9cm]

{Anas Shatnawi\footnote{anasshatnawi@gmail.com}, Hafedh Mili\footnote{mili.hafedh@uqam.ca}, Manel Abdellatif, Ghizlane El Boussaidi, \\Yann-Ga\"el Gu\'eh\'eneuc, Naouel Moha, Jean Privat}\\[0.5cm]

\textit{LATECE Technical Report 2017-1, LATECE Laboratoire, Universit\'e du Qu\'ebec \`a Montr\'eal, Canada}

\end{center}

\begin{abstract}
Java Enterprise Edition (JEE) applications are implemented in terms of a set of components developed based on several JEE technologies including, but not limited to, Servlet, JSP, JSF, EJB, JavaBeans. These JEE technologies rely on a varied set of communication mechanisms to communicate between each others. Examples of these communication mechanisms are HTTP requests, Remote Method Invocation (RMI), Java DateBase Connectivity (JDBC), etc. These communication mechanisms represent program dependencies between JEE components.
However, one communication mechanism can be implemented following different implementation ways by different JEE technologies. Therefore, to be able to detect related dependencies, we identify these implementation ways used by a set of JEE technologies. In this technical report, we focus on the Web tier technologies that are Servlets, JSPs and JSFs. Also, we present how these technologies access the JavaBeans and Manage Beans components.
\end{abstract}

\section{Introduction}
\label{Introduction}

\begin{figure*}[!h]
	\begin{center}
		\includegraphics[width=0.5\textwidth]{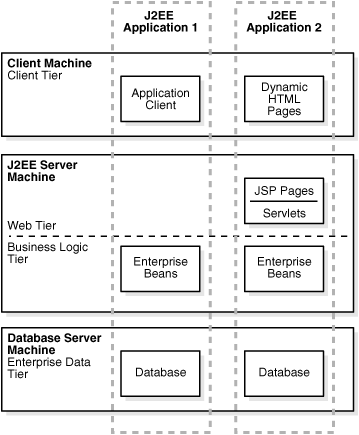}
		\caption{Architecture of multi-tiered JEE applications}
		\label{fig:multitierArch}
	\end{center}
\end{figure*}

Java Enterprise Edition (JEE) allows to manage, develop and deploy web applications based on the multi-tiered distributed model.
In this context, the application logic of a web application is decomposed into a set of components distributed among different tiers based on their functionalities. 
These components can be deployed on different machines. This is based on the tier that each component belongs. Figure \ref{fig:multitierArch} shows an example of two web applications. These applications are split into four tiers;  client, web, business logic and data tiers. 
The client tier can be of two types. The first one denotes to application clients that are installed on client machines and access business logic functionalities running on server machines. The second type refers to HTML pages that use HTTP requests to communicate with web tier components.
The web tier implements the presentation logic of the application based on Servlets, JavaServer Pages (JSPs) and JavaServer Faces (JSFs).
The business tier encapsulates the business logic of an application using Enterprise Beans.
The data tier stores the enterprise data.

The JEE technologies rely on a varied set of communication mechanisms to communicate between each others. Examples of these communication mechanisms are HTTP requests, Remote Method Invocation (RMI), Java DateBase Connectivity (JDBC), etc. These communication mechanisms represent program dependencies between JEE components.

However, one communication mechanism can be implemented following different implementation ways by different JEE technologies. Therefore, to be able to detect related dependencies, we identify these implementation ways used by a set of JEE technologies. In this technical report, we focus on the Web tier technologies that are Servlets, JSPs and JSFs. Also, we present how these technologies access the JavaBeans and Manage Beans components. To do so, we studied on the Oracle’s JEE specification to identify the communication mechanisms.

The rest of this report is organized as follows. First, we present a background about  brief introduction about Servlet, JSP, JSF, JavaBeans and Manage Beans JEE technologies in Section \ref{sec:JEETech}. 
Next, we show how server containers configure the URL of HTTP Requests in Section \ref{sec:URLCong}. Then, Section \ref{sec:interactionSerJSPJSF} discusses the interaction dependencies that can be existed between Servlets, JSPs and JSFs. Section \ref{sec:accessBeans} shows how to access JavaBeans and Manage Beans components at JSP and JSF pages.

\section{Anatomy of Web Tier Technologies in JEE}
\label{sec:JEETech}
In this section, we discuss the background needed to understand the main concepts of: (1) Servlets, (2) JSPs, (3) JSFs, (4) JavaBeans and (4) Managed Beans technologies.

\subsection{Servlets}

\subsubsection{What are Servlets}
Servlets are server-side components that allow one to extend the web server capabilities to offer functionalities for client requests (e.g., HTTP requests). They represent the gateway between clients (e.g., Web browsers) and servers.
Servlets are implemented based on object-oriented Java code. This is based on a set of interfaces and classes located at the \textit{javax.servlet} and \textit{javax.servlet.http} packages. The Java implementation of Servlets allows them to have the advantages of other Java-based technologies (e.g., JDBC, RMI, EJB).

\subsubsection{Life Cycle of Servlets}
\begin{figure}[h]
	\begin{center}
		\includegraphics[width=\textwidth]{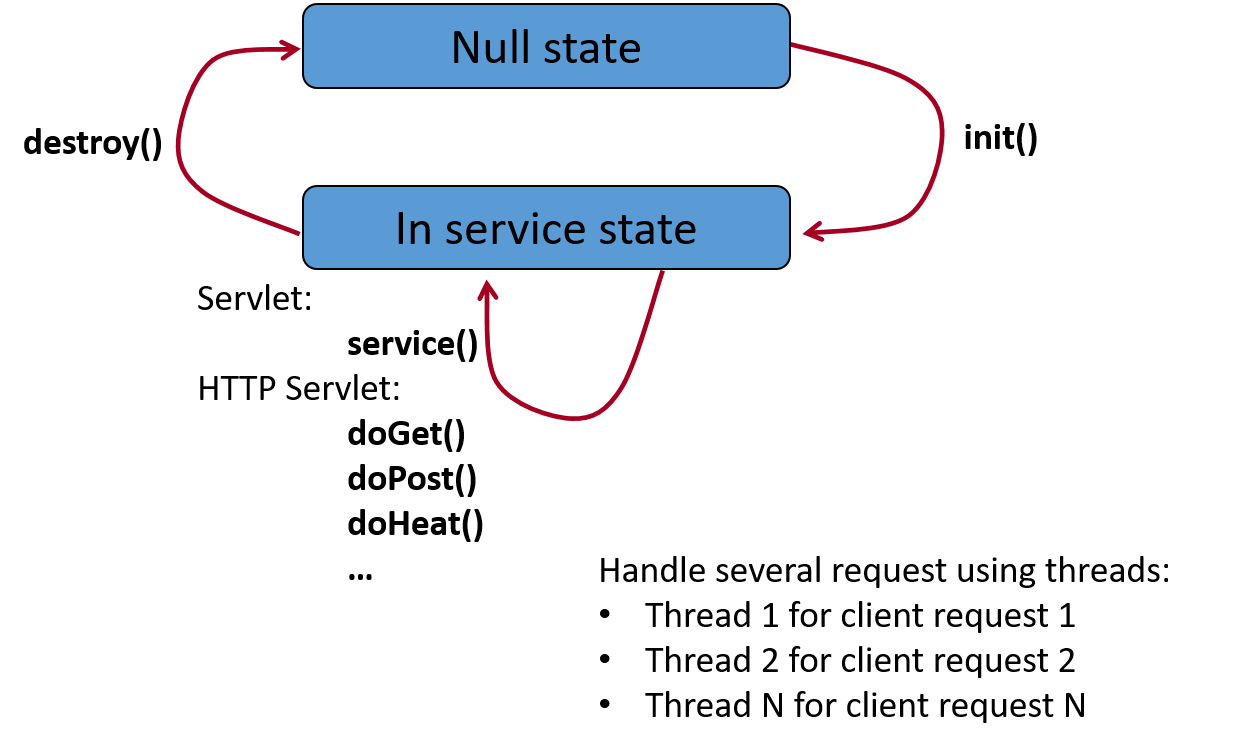}
		\caption{Life cycle of a Servlet}
		\label{fig:servlet-life-cycle}
	\end{center}
\end{figure}

The life cycle of a Servlet starts at the time that the Web container receives a request of a relative-URL corresponding to this Servlet. The Web container automatically initializes the Servlet by invoking the \textit{init()} method. This is only done for the first request (not getting called for next requests). Then, each time the Web container receives a new request, it calls the \textit{service() method} (using a new thread) and passing the \textit{request} and \textit{response} objects to the Servlet. 
The request object attaches information related to the client (e.g., request parameters). The response one is used to store the results that the Servlet wants to return to the client. 
At the end of the servlet life cycle, the Web container calls the \textit{destroy()} method. This throws the Servlet for the Java garbage collection. Figure \ref{fig:servlet-life-cycle} shows the state diagram of the life cycle of a Servlet.

\subsubsection{Types of Servlets}
There are two types of servlets:  \textit{generic} and \textit{HTTP} Servlets. 

\begin{enumerate}
\item \textbf{Generic Servlet} is used for general requests. It is implemented by extending the \textit{javax.servlet.GenericServlet} class. A generic Servlet follows the default Servlet life cycle (\textit{init()}, \textit{service()} and \textit{destroy()}).

\item \textbf{HTTP Servlet} is a subclass of the generic Servlet that is only specific for handling HTTP requests. It is implemented by extending the \textit{javax.servlet.HTTP.HTTPServlet} class. This type overrides the default Servlet life cycle to be suitable for HTTP requests. To this end, the \textit{doPost()}, \textit{doPut} and \textit{doGet()} methods are added to respectively handle post, put and get HTTP requests. These methods overdrive the \textit{service()} method. An example of an HTTP Servlet is presented in Listing \ref{servletExample}.
\end{enumerate}

\begin{lstlisting}[caption=An example of HTTP servlet, label=servletExample]

// Import required java libraries
import java.io.*;
import javax.servlet.*;
import javax.servlet.http.*;

// Extend HttpServlet class
public class ServletExample extends HttpServlet {

	public void init() throws ServletException {
		// Initialization code
	}
	
	public void service(ServletRequest request, ServletResponse response) 
	throws ServletException, IOException{
		// Servlet code
	}
	
	public void doGet(HttpServletRequest request, HttpServletResponse response)
	throws ServletException, IOException {
		// Servlet code
	}
	
	public void doPost(HttpServletRequest request, HttpServletResponse response)
	throws ServletException, IOException {
		// Servlet code
	}
	
	public void destroy() {
		// Finalization code
	}
}

\end{lstlisting}

\subsection{JSP}

\subsubsection{What are JSPs}
JavaServer Pages (JSP) is a server-side scripting language that is used to implement Java Servlet with respect to support the development of dynamic web pages that represent the presentation layer of Web applications. 

A JSP page is an HTML page with special tags that is structured as a XML-based text file. Usually, the text file has \textit{.jsp} extension. A JSP page could consist of a mixed of static content (like HTML and XML) and dynamic content (dynamic JSP tags). In addition, inside the JSP page, developers are allowed to embed Java code fragments and to invoke external Java components such as JavaBeans.
To this end, JSP supplies a collection of scripting elements in the form of XML. For example, JSP scriptlet (\textit{<\% Java code \%>}) enables to insert Java code. 

\subsubsection{Life Cycle of JSPs}
Before the running of a JSP, the Web container translated the JSP page into an equivalent Java Servlet that processes client requests like a normal Servlet. Therefore, JSP pages have the same life cycle of Servlets.
The translation of JSP pages is as follows. Once a web container receives a URL related a JSP page (\textit{http://host/myPage.jsp}), it calls the JSP translator that generates an equivalent \textit{.java} file (\textit{myPage.java}).
Then, it is compiled to generate an executable file (\textit{myPage.class}).

\subsection{JSF}
\subsubsection{What are JSFs}

Similar to JSPs, JSFs are also built on the top of Java Servlet to provide a component framework for developing server-side Java Web applications. JSFs provide a set of reusable  components via their tag libraries. 
JSFs are implemented based on Facelets technology which allows to develop view presentation through HTML style templates that are organized in terms of a tree of components. This means that it supports to reuse its components through the templating and composition of reusable components in other Java based view technologies such as JSP pages. 

JSF is consists of two main parts. The first one is composed of APIs that represent components that can be used for server-side validation, page navigation, handling events, etc. The second part consists of a collection of tag libraries that provide access to add the reusable components to a Web page based on XML namespace declarations. JSFs are usually developed using XHTM pages (\textit{.xhtml} extension).

\subsubsection{JSF Tag Libraries}
JSF components are mainly organized in five tag libraries. Such that each library specializes for related functionalities. These are HTML, Facelets, Core, JSTL Core and JSTL Functions tag libraries.
The HTML tag library consists of component tags related to user interface objects, e.g., \textit{<h:body>}, \textit{<h:outputText>}, \textit{<h:inputText>}, etc. Tag libraries should be imported before any use. This is based on a relative URL corresponding to each tag library (e.g., \textit{http://java.sun.com/jsf/html} corresponds to the HTML tag library). Listing \ref{JSFStructure} shows an example of a JSF page that imports the HTML tag library. It reuses two HTML components; \textit{<h:head>} and \textit{<h:body>}.

\begin{lstlisting}[caption=JSF page structure, label=JSFStructure]

<html lang="en" xmlns="http://www.w3.org/1999/xhtml"
xmlns:h="http://java.sun.com/jsf/html">
	<h:head>
		<title>Page Title</title>
	</h:head>
	<h:body>
		// body code
	</h:body>
</html>
\end{lstlisting}

\subsection{JavaBeans}
JavaBeans is a component model that allows to implement reusable components. These components can be used without the need to understand their inner implementations. Programmatically, a JavaBeans component is coded based on a normal Java class with respect to special three characteristics. Such characteristics are that: 
\begin{enumerate}
\item the class should be serializable (to save the state of an object), 

\item has a no-argument contractor (to instantiate the object), and

\item supports public setter and getter method properties (to get and set the values of private variables). 
\end{enumerate}

Listing \ref{javaBeans} shows an example of a JavaBeans component that encapsulates student information.

\begin{lstlisting}[caption=Example of a JavaBeans component, label=javaBeans]
public class Student implements java.io.Serializable {
	
	private String name;
	private int ID;
	...
	
	public Student (){
	}
	
	public String getName() {
		return this.name;
	}
	
	public int getId() {
	return this.id;
	}
	
	public void setName(String name) {
	this.name = name;
	}
	
	public void setId(int id) {
		this.id = id;
	}
	...
}
\end{lstlisting}

\subsection{Managed Beans}


It is a regular JavaBeans component model that is designed for JSF technology. Thus, it also allows one to implement components based on Java classes that are created with no argument constructors, a set of setter and getter properties, and a set of methods performing services for the components. 
In JSF, Managed Beans components can be used to validate data, to handle events, to process data, etc.  
However, they need to be registered at the JSF configuration file before reuse. This can be done based on two mechanisms; configuration files and annotations.

\subsubsection{Configuration of Managed Beans Using Configuration File: the \textit{filefaces-config.xml} file}
Listing \ref{registerMbeansinJSF} shows an example of how to register a Managed Beans at the \textit{filefaces-config.xml} configuration file.

\begin{lstlisting}[caption=Example of Managed Beans registration at JSF configuration file, label=registerMbeansinJSF]
<managed-bean>
	<managed-bean-name>YouCanUseME</managed-bean-name>
	<managed-bean-class>myPackage.MyBean</managed-bean-class>
	...
</managed-bean>
\end{lstlisting}

\subsubsection{Configuration of Managed Beans Using Annotations}
Starting from the JSF 2.0 specification, the configuration of Managed Beans components can be performed based on Bean annotations that make the process easier to be managed. 
The annotations are embedded in the implementation code to declare a set of attributes that a Managed Bean supports. Such annotations are used for different reasons. For example, the \textit{@ManagedBean(name=“idBean")} is used to declare a Beans component. In addition, the annotations enable to declare a reference name for an attribute of a Managed Beans. This is based on \textit{@ManagedProperty(value="\#{referenceName}")}.

Listing \ref{registerMbeansAnnotations} presents an example that uses annotations to register a Managed Beans component and a declaration of a proprieties  of this Beans.

\begin{lstlisting}[caption=Example of Managed Beans registration using annotaions, label=registerMbeansAnnotations]

@ManagedBean(name = "YouCanUseME")
public class MyBean {

	@ManagedProperty(value="#{message}")
	private Message message;
	...
}

\end{lstlisting}

\section{Configuration of Relative URLs of Server Pages}
\label{sec:URLCong}
In the context of JEE Web applications, clients use HTPP requests to demand services. These requests are structured as Web addresses that reference Web pages through relative-URLs. The requests are handled via dynamic/static Web pages that can be implemented based on different technologies such as Servlet, JSP, HTML, JSF, etc. 

\begin{figure*}[!h]
	\begin{center}
		\includegraphics[width=\textwidth]{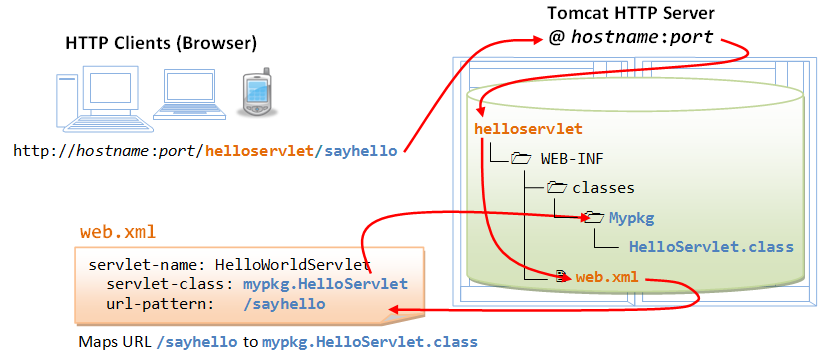}
		\caption{Mapping URL to a Servlet}
		\label{fig:urlConfig}
	\end{center}
\end{figure*}

Once the web container receives an HTTP request, it looks in deployment descriptors (e.g., \textit{web.xml}) to identify the corresponding handler Web page related to the requested URL. Figure \ref{fig:urlConfig} shows the process of URL mapping performed by the Web container. There are two ways of defining URLs for Web pages; the \textit{web.xml} deployment descriptor and annotations.

\subsection{Servlet Declaration at Deployment Descriptor (\textit{web.xml})}
The \textit{web.xml} file is an XML-format file located in the \textit{WEB-INF/} directory of JEE applications. 
In this XML file, there two main particular tags that are used by the Web container to identify the server pages corresponding to the requested URLs. These are the \textit{<servlet>} and the \textit{<servlet-mapping>} tags. To better understand these tags, we present an illustrative example that is presented in Listing~\ref{web:xml}.

The \textit{<servlet>} tag is used to declare a server page. 
The declaration consists of three main elements. The first one is \textit{<servlet-name>} which assigns an identity name used to refer to the server page. In our example, \textit{name1} is given to the Servlet and \textit{name2} is given to the JSP page. 
The second element is used to determine the implementation Java file that executes the Servlet. This can be of two sub-elements based on the type of the executable Servlet. \textit{<servlet-class>} refers to a Servlet class. In our example, \textit{com.jee.MyFirstServlet}, and \textit{<jsp-file>} refers to a JSP page (e.g., \textit{Page1.jsp}). 
The third element (\textit{<init-param>}) is used to pass initial parameters to the server page. In our example, the Servlet will receive a parameter called \textit{ParameterName} that has \textit{ParameterValue} as a value.

The <servlet-mapping> tag maps requests that confirm to a URL pattern to one of the already defined server pages. The URL pattern is defined based on the \textit{<url-pattern>} element. In our example, \textit{com.jee.MyFirstServlet} is the handler of the \textit{"/ServletURL"} URL pattern, \textit{Page1.jsp} is the handler of three URL patterns. These patterns are \textit{"/myJSPPage.JSP"}, \textit{"/myHTMLPage.html"} and \textit{"/hi"}, \textit{Page2.JSP} is the handler of any URL ended by \textit{.jsp} (URL patter is \textit{*.JSP}).

A URL pattern can be (\textit{*}), such that the corresponding server page is the handler of any requested URL. For example, the \textit{com.jee.MySecondServlet} will be called to handle any request.

\begin{lstlisting}[caption=An example of URL configuration in \textit{web.xml} file, label=web:xml]
<web-app xmlns="http://java.sun.com/xml/ns/javaee" version="2.5">
	<servlet>
			<servlet-name>name1</servlet-name>
			<servlet-class>com.jee.MyFirstServlet</servlet-class>
				<init-param>
						<param-name>ParameterName</param-name>
						<param-value>ParameterValue</param-value>
				</init-param>
			<servlet-name>name2</servlet-name>
			<jsp-file>Page1.jsp</jsp-file>
			<servlet-name>name3</servlet-name>
			<jsp-file>Page2.JSP</jsp-file>
			<servlet-name>name3</servlet-name>
			<jsp-file>com.jee.MySecondServlet</jsp-file>
	</servlet>
	
	<servlet-mapping>
			<servlet-name>name1</servlet-name>
			<url-pattern>/ServletURL</url-pattern>	
			<servlet-name>name2</servlet-name>
			<url-pattern>/myJSPPage.JSP</url-pattern>			
			<servlet-name>name2</servlet-name>
			<url-pattern>/myHTMLPage.html</url-pattern>	
			<servlet-name>name2</servlet-name>
			<url-pattern>/hi</url-pattern>
			<servlet-name>name3</servlet-name>
			<url-pattern>*.JSP</url-pattern>
			<servlet-name>name4</servlet-name>
			<url-pattern>/*</url-pattern>
			
	</servlet-mapping>	
</web-app>
\end{lstlisting}

\subsection{Servlet Declaration at (\textit{@WebServlet} Annotation)}

Starting form the Servlet 3.0 specification, servlet declaration is not required to be done at the \textit{web.xml} deployment descriptor. Instead, it can be done based on the \textit{@WebServlet} annotation. Listing~\ref{annotation} shows examples of servlet declaration where the first one maps \textit{MyFirstServlet} to the \textit{"/ServletURL"} URL pattern and the second one maps MySecondServlet to a set of three URL patterns; \textit{"/myHTMLPage.html"}, \textit{"/myJSPPage.jsp"} and \textit{"/myJSFPage.xhtml"}.

\begin{lstlisting}[caption=Examples of URL configuration using \textit{@WebServlet} annotation, label=annotation]
// One URL pattern
@WebServlet("/ServletURL")
public class MyFirstServlet extends HttpServlet
{
	...
}

// More than one URL patterns 
@WebServlet(urlPatterns = {"/myHTMLPage.html", "/myJSPPage.jsp", "/myJSFPage.xhtml"})
public class MySecondServlet extends HttpServlet{
	...
}

\end{lstlisting}

\section{Communication Mechanisms Among Server Pages: Servlet, JSP and JSF}
\label{sec:interactionSerJSPJSF}

Servlets, JSPs and JSFs can forward/include output to another server pages. 
In this section, we present the different communication mechanisms between Servlets, JSPs and JSFs.

\subsection{Communication Mechanisms of Servlet Server Pages}

Servlets invoke other server pages based on a stander interface called \textit{javax.servlet.RequestDispatcher}. This interface is obtained from the \textit{servlet context} instance \textit{javax.servlet.ServletContext} that allows the communication between Servlets and their Web container. Each web container has only one \textit{ServletContext} provided by the Java Virtual Machine. 
Obtaining the instance of this servlet context from the Servlet instance can be done through the \textit{this.getServletContext()} method. 

This servlet context instance provides access to the instance of the request dispatcher corresponding to a given relative-URL related to the target server page through the \textit{getRequestDispatcher()} method. The URL is given as a parameter (e.g., string) to this method (e.g., \textit{getRequestDispatcher("/jsp/mypage.jsp");}). 
Then, there are two ways to call an URL. The first one is to include the output of other Servlet. This uses the \textit{include()} method. The second way is to handle interaction control to another Servlet. This uses the \textit{forward()} method. These methods take instances of HTTP requests and responses as arguments.

Listing~\ref{servletInvokeURL} presents examples of how \textit{MyFirstServlet} invokes \textit{MySecondServlet} based on three calling ways.


\begin{lstlisting}[caption=Inoking a URL, label=servletInvokeURL]
@WebServlet("/ServletURL")
public class MyFirstServlet extends HttpServlet
{
		...
		// 1)Using include
		RequestDispatcher dispatcher =
		getServletContext().getRequestDispatcher("/MySecondServlet");
		dispatcher.include(request, response);
		...
		// 2)Using forward
		RequestDispatcher dispatcher =
		getServletContext().getRequestDispatcher("/MySecondServlet");
		dispatcher.forward(request, response);
		...
		// 2)Using another scenario
		getServletContext().getRequestDispatcher("/MySecondServlet").forward(request, response);
		...
}

\end{lstlisting}

\subsection{Communication Mechanisms JSP Server Pages}

JSPs invoke server pages based on their URLs using seven different programmatical ways. These are: (1) the HTML form, (2) the \textit{jsp:include} action tag , (3) the \textit{include} directive tag, (4) the \textit{jsp:forward} action tag, (5) the stander JSTL taglib directive (Core tag library), (6) using normal Java code, and (7) the \textit{page} directive tag. 

In the remaining of this section, we explain these ways.

\subsubsection{The HTML Form}

	It is the traditional HTML form that is used to create a form for the user input. It is composed of the \textit{<form>} tag that has a set of attributes to parameterize the form. Such attributes are \textit{action} and \textit{method}. 
	The \textit{action} attribute determines where to send the form based on a relative-URL that corresponds to the target server page that will handle the form. The \textit{method} attribute defines the HTML method. It can be of different types (i.e., \textit{get}, \textit{post}, \textit{put}). This determines which method(s) will be invoked in the handler server page (e.g., service methods in the Servlets).
	
	Moreover, the \textit{<form>} tag could include other elements like \textit{<input>}, \textit{<button>}, \textit{<label>}, etc. For example, <input> defines a field that allows the users to input their data.

Listing~\ref{htmlForm} shows an example of an HTML form that contains two input fields and one button. This form will be sent to \textit{myPage.jsp} to be handled.
	
	\begin{lstlisting}[caption=JSP page inokes a URL using a HTML form, label=htmlForm]
	...
	<form action="/myPage.jsp" method="get">
		 First name: <input type="text" name="fname"><br>
		 Last name: <input type="text" name="lname"><br>
		 <input type="submit" value="Submit">
	</form> 
	...
	\end{lstlisting}

\subsubsection{The \textit{jsp:include} Action Tag}
	
	It is an action tag that JSP pages use to dynamically include the content of another server page. It is codified based on the \textit{<jsp:include>} tag where a relative-URL of the target server page is given using the \textit{file} attribute.

Listing~\ref{includeActionTag} presents an example of a JSP page that includes the content of \textit{myPage.jsp} and \textit{myServlet}.
	
	\begin{lstlisting}[caption=JSP page inokes a URL using \textit{jsp:include} action tag, label=includeActionTag]
	...
	<jsp:include page="/myPage.jsp." flush="true" />
	...
	<jsp:include page="/myServlet" flush="true" />
	...
	\end{lstlisting}
	
\subsubsection{The \textit{include} Directive Tag}
	It is used to insert the content of another server page in the current JSP page at the translation time. The \textit{include} directive is codified using two tags. These are the \textit{<\%@ include\%>} and \textit{<jsp:directive.include>} tags. The relative-URL is given as a parameter value of the \textit{file} attribute.
    
	Listing~\ref{includeDirective} shows examples of the two ways that the include directive can be codified.
	
	\begin{lstlisting}[caption=Example of  \textit{include} in JSP, label=includeDirective]
	...
	<%@ include file="/myPage.jsp" %> 
	...
	<jsp:directive.include file="/myPage.jsp" />
	...
	\end{lstlisting}

\subsubsection{The \textit{jsp:forward} Action Tag}
	This action tag is used to forward HTTP requests to other server pages. It is exactly equal to the request forward way used in the Servlet. Pragmatically, it is based on the \textit{<jsp:forward>} action tag, where the target server page is determined based on a relative-URL attached to the \textit{page} attribute. Furthermore, parameters can be optionally attached to the request. This is based the \textit{<jsp:param>} action tag that is putted as a child of \textit{<jsp:forward>}.
	
	Listing~\ref{JSPtoServlet:forward} presents an example of a JSP page that forwards a request to \textit{myPage.jsp} based on \textit{<jsp:forward>}. In this example, we firstly show how \textit{<jsp:forward>} can be used without passing parameters. Then, we present how to pass two parameters in the forwarded request using the <jsp:param> tag. These parameters are \textit{name} and \textit{id} that respectively have \textit{"sami"} and \textit{"124"} as values.

\begin{lstlisting}[caption=Example of \textit{jsp:forward} in JSP, label=JSPtoServlet:forward]
	...
	// Withour parameters
	<jsp:forward page="myPage.jsp" />
	
	//With parameters
	<jsp:forward page="myPage.jsp"> 
		<jsp:param name="name" value="Sami" /> 
		<jsp:param name="id" value="123" /> 
	</jsp:forward>
	...	
	\end{lstlisting}
	
\subsubsection{The Stander JSTL taglib Directive: Core Tag Library}	
The JSP Stander Tag Library (JSTL) is a set of reusable tags (i.e., components) that provide pre-implemented functionalities. These reusable tags are clusters based on their provided functionalities into groups. One of these tag libraries is the \textit{core} one that consists of a set of basic operations frequently used by software developers.
    
Before reusing the core tag library, it is necessary to declare that inside the JSP implementation. To do so, the \textit{<\%@ taglib prefix="c" uri="http://java.sun.com/jsp/jstl/core" \%>} tag should be coded such that \textit{c} is the tag identity. 

There are two interested tags that can be used to connect the JSP page with other server pages. The first one is the \textit{<c:redirect url="relative-URL"/>} tag that redirects the request to another server page related to the given relative URL. The second one is the \textit{<c:url>} tag that creates a URL.
	
Moreover, both tags (\textit{<c:redirect>} and \textit{<c:url>}) accept to pass parameters to the target server page. This is based on the \textit{<c:param>} tag, where the \textit{name} attribute determines the parameter name and the \textit{value} one gives the parameter value. 

Listing~\ref{JSPtoServlet:JSTL} presents an example of a JSP code that firstly declares the usage of the core tag library. Then, it redirects the request to \textit{myPage.jsp} with and without parameters respectively. Finally, it creates a URL to \textit{myPage.jsp} with two parameters.

	\begin{lstlisting}[caption=Example of core tag library in JSP, label=JSPtoServlet:JSTL]
	...
	// To declare the use of the core tags
	<%@ taglib prefix="c" uri="http://java.sun.com/jsp/jstl/core" %>
	// To redirect the request to myPage.jsp without any parameter
	<c:redirect url="/myPage.jsp"/>
	// To redirect the request to myPage.jsp with two parameters
	<c:redirect url="/myPage.jsp">
		<c:param name="name" value="sami"/>
		<c:param name="id" value="123"/>
	<c:redirect>
	// To create a URL with two parameters
	<c:url value="/myPage.jsp" var="completeURL">
		<c:param name="name" value="sami"/>
		<c:param name="id" value="123"/>
	</c:url>
	...
	\end{lstlisting}
    
\subsubsection{Using Normal Java Code}
	JSP pages can use the same way that Servlets use to forward or include HTTP requests to other server pages. This based on embedding Java code inside the JSP script using the JSP \textit{scriptlet} tags. These \textit{scriptlet} tags can be of two equal forms; (1) \textit{<\% ... \%>} and (2) \textit{<jsp:scriptlet> ... </jsp:scriptlet>}.

Listing~\ref{JSPtoServlet:Code} presents an example of a JSP code that forwards the request to \textit{myPage.jsp} based on the JSP \textit{scriptlet} tags.
	
	\begin{lstlisting}[caption=Example of Java code embedded in JSP to forward request, label=JSPtoServlet:Code]
	...
	// To include a request
	<% 	RequestDispatcher dispatcher =
	getServletContext().getRequestDispatcher("/myPage.jsp");
	dispatcher.include(request, response);
	%>
	// To forward a request
	<jsp:scriptlet> RequestDispatcher dispatcher =
	getServletContext().getRequestDispatcher("/myPage.jsp");
	dispatcher.forward(request, response);
	</jsp:scriptlet>
	...
	\end{lstlisting}	
	
\subsubsection{The \textit{page} Directive Tag}
	
	JSP pages can inform the Web container which server page to call in the case of exceptions at runtime. This can be done through the \textit{errorPage} attribute under the \textit{page} directive tag. 

The \textit{page} directive tag can be codified using two tag elements; \textit{<\%@ page\%>} and \textit{<jsp:directive.page>}. The relative URL of the handler server page is given as a value for the  \textit{errorPage} attribute (i.e., \textit{errorPage="relative-URL"}). 

	Listing~\ref{JSPtoServlet:page} presents an example of a JSP code that calls \textit{errorPage.jsp} in the case of errors based on the two forms of the \textit{page} directive tag.
		
	\begin{lstlisting}[caption=Example of the \textit{page} directive tag in JSP, label=JSPtoServlet:page]
	...
	// First form
	<%@ page errorPage="errorPage.jsp" %>
	// Second form
	<jsp:directive.page errorPage="errorPage.jsp"/>
	...
	\end{lstlisting}

\subsection{Communication Mechanisms of JSF Server Pages}
As it has been mentioned before, JSF is built based on a set of reusable components organized in a collection of tag libraries. Some of these components can be used to interact with server pages. We identify three tags (components) that are: (1) \textit{commandButton}, (2) \textit{commandLink}, and (3) \textit{href}.

\subsubsection{The \textit{commandButton} Tag}	
	The \textit{commandButton} tag implements a submit button. It is implemented under the HTML tag library. Thus, the corresponding import statement should be  codified before the use of this tag. This is done through the \textit{<html xmlns:h="http://java.sun.com/jsf/html">} import statement that allows to access the HTML tag reference through the \textit{h} identity. 
	
	The codification of the \textit{commandButton} tag is based on \textit{<h:commandButton>}. The relative URL is given as a value of the \textit{action} attribute.

Listing~\ref{JSF:commandButton} presents an example that imports the HTML tag reference, then it connects the \textit{<h:commandButton>} element to \textit{myPage.jsp}.

\begin{lstlisting}[caption=Example of \textit{commandButton} in JSF, label=JSF:commandButton]
	<html xmlns="http://www.w3.org/1999/xhtml"
		xmlns:h="http://java.sun.com/jsf/html">
		...
		<h:commandButton id="submit" value="Submit" action="/myPage.jsp"/>
		...
\end{lstlisting}
	
\subsubsection{The \textit{commandLink} Tag}
It is another JSF component implemented under the HTML tag library. It is used to create a \textit{hyperlink} to another server page. The codification is based on the \textit{<h:commandLink>} element. The relative URL is given as a value of the \textit{action} attribute. Listing~\ref{JSF-commandLink} presents an example that imports the HTML tag reference, then it connects the \textit{<h:commandLink>} element to \textit{myPage.jsp}.
	
\begin{lstlisting}[caption=Example of \textit{commandLink} in JSF, label=JSF-commandLink]
	<html xmlns="http://www.w3.org/1999/xhtml"
		xmlns:h="http://java.sun.com/jsf/html">
		...
		<h:commandLink value="LINk" action="/myPage.jsp"/>
		...
\end{lstlisting}
	
\subsubsection{The \textit{href} Tag}
It is also an HTML component that is used to create a \textit{hyperlink} to another page, but it does not generate any action event. 
Another difference compared to the above hyperlink action is that the \textit{commandLink} could accept to indicate the name of the page without the extension (\textit{myPage}), while \textit{href} requires the extension (\textit{myPage.xhtml}).	

The codification is based on the \textit{<a>} element. The relative URL is given as a value of the \textit{href} attribute (see Listing~\ref{JSF:href}).
	
\begin{lstlisting}[caption=Example of \textit{href} in JSF, label=JSF:href]
	...
	<a href="page.xhtml">Message</a> 
	...
\end{lstlisting}

\section{Referencing JavaBeans and Manage Beans Components}
\label{sec:accessBeans}

JSPs and JSFs use Beans components, e.g., \textit{JavaBeans} and \textit{Managed Beans} to access business tier functionalities. This can be done through two mechanisms provided by JSP and JSF technologies. These mechanisms are: (1) the JSP tag libraries and (2) the {expression language}.

\subsection{Access Beans Based on Tag Libraries}
Beans components can be referenced using the JSP \textit{useBean} action tag that is used to declare the use of a JavaBean component in a JSP page. This deceleration allows one to reuse the beans in terms of a scripting variable accessed by other scripting elements used in the JSP. 

Together with the \textit{<jsp:useBean>} action tag, it is possible to use the \textit{<jsp:getProperty/>} and \textit{<jsp:setProperty/>} action tags. The \textit{<jsp:getProperty/>} is used to get an access to the getter methods. The \textit{<jsp:setProperty/>} refers to the access to the setter methods. 

Listing~\ref{AccessBeans:jspTag} shows the syntax of the \textit{useBean}, \textit{<jsp:getProperty/>} and <jsp:setProperty/> action tags.. In this syntax, the \textit{id} attribute is used to give a unique identifier name. 
The \textit{class} attribute determines the Java class that has the implementation of the Beans. The \textit{scope} attribute is to specify the scope of the Beans that could be of four types; a page, a request, a session or an application.
For the \textit{<jsp:getProperty/>} and \textit{<jsp:setProperty/>} action tags, 
the \textit{name} attribute refers to the Beans's \textit{id} that confirms to the \textit{useBean} action one. The \textit{property} attribute is the setter and getter method names that should be called. The \textit{value} attribute is the parameter of the setter method. 

\begin{lstlisting}[caption=The syntax of \textit{useBenas} tag, label=AccessBeans:jspTag]
...
<jsp:useBean id = "bean's id" class = "bean's class" scope = "bean's scope">
   <jsp:setProperty name = "bean's id" property = "property name"  
      value = "value"/>
   <jsp:getProperty name = "bean's id" property = "property name"/>
   ...........
</jsp:useBean>...
\end{lstlisting}

\subsection{Expression Language to Access Beans Components}

The expression should be between the  \textit{ \$\{}  and \textit{\}} characters (i.e., \textit{ \$\{expression\}}). The expression itself can be composed of a combination of several components. Such components are: (1) mathematical expressions, (2) logical expressions, and (3) references to Beans (classes, methods, and attributes).

The reference of a Beans component is directly though its name \textit{ \$\{beans's name\}}. The reference to its attributes and methods is based on the \textit{.} concatenation of the beans's name and the name of the desired attribute/method (i.e., \textit{ \$\{beans's name.attribute's name\}} and \textit{ \$\{beans's name.method's name\}}).

Expression language can be used in two main contexts: (1) inside template text and (2) the value of a tag attribute that can accept runtime expressions.

Listing~\ref{AccessBeans:EL} shows three examples of accessing Beans components following three different ways.

For more detail about the Expression Language, please refer to \cite{EL}.

\begin{lstlisting}[caption=Examples of accessing Beans using expression language, label=AccessBeans:EL]
...
// Inside text
Hello ${student.firstName}, how are you?

//Without passing a parameter
<h:commandButton action="#{trader.buy}" value="buy"/>

//Passing a parameter 
<h:commandButton action="#{trader.buy('SOMESTOCK')}" value="buy"/> 
...
\end{lstlisting}

\section{Conclusion}
JEE applications are composed of a set of components implemented using different technologies. These components rely on a variety of mechanisms to communicate among each others. In this technical reports, we discussed the set of implementation mechanisms that are used at the Web tier of JEE applications. 
We depended on the Oracle’s JEE specification to identify these communication mechanisms.
We covered mechanisms used by components implementing using Servlet, JSP, JSP, JavaBeans and Managed Beans technologies. 

As future directions, we want to develop a static analysis tool that is able to detect all of these dependencies and build a dependency call graph. This call graph will allow us to understand relationships among the different components which is required for several software engineering tasks. Also, we want to cover other JEE tier and technologies. 

\bibliographystyle{unsrt}
\bibliography{main}  

\end{document}